\documentclass[aps,pra,twocolumn,superscriptaddress,showpacs,nofootinbib]{revtex4-1}

\usepackage[english]{babel}
\usepackage{blindtext}
\usepackage{graphicx}
\usepackage{subfigure}
\usepackage{amsmath,color}
\def\gappeq{\mathrel{ \rlap{\raise.5ex\hbox{$>$}}
                      {\lower.5ex\hbox{$\sim$}} } }
\def\lappeq{\mathrel{ \rlap{\raise.5ex\hbox{$<$}}
                      {\lower.5ex\hbox{$\sim$}} } }

\begin{document}

\title{Anisotropic and long-range vortex interactions in two-dimensional dipolar Bose gases}

\author{B. C. Mulkerin} 
\affiliation{School of Physics, University of Melbourne, Victoria 3010, Australia}
\author{R. M. W. van Bijnen}
\affiliation{Eindhoven University of Technology, PO Box 513, 5600 MB Eindhoven, The Netherlands}
\author{D. H. J. O'Dell}
\affiliation{Department of Physics and Astronomy, McMaster University, Hamilton, Ontario, L8S 4M1, Canada}
\author{A. M. Martin}
\affiliation{School of Physics, University of Melbourne, Victoria 3010, Australia}
\author{N. G. Parker}
\affiliation{Joint Quantum Centre Durham--Newcastle, School of Mathematics and Statistics, Newcastle University, Newcastle upon Tyne, NE1 7RU, United Kingdom.}


\begin{abstract}

We perform a theoretical study into how dipole-dipole interactions modify the properties of superfluid vortices within the context of a two-dimensional atomic Bose gas of co-oriented dipoles.  The reduced density at a vortex acts like a giant anti-dipole, changing the density profile and generating an effective dipolar potential centred at the vortex core whose most slowly decaying terms go as   $1/\rho^2$ and $\ln(\rho)/\rho^3$.   These effects modify the vortex-vortex interaction which, in particular, becomes anisotropic for dipoles polarized in the plane.  Striking modifications to vortex-vortex dynamics are demonstrated, i.e.\ anisotropic co-rotation dynamics and the suppression of vortex annihilation. 

\end{abstract}

\maketitle

Ferrohydrodynamics (FHD) describes the motion of fluids made of particles with magnetic (or electric) dipoles \cite{rosensweig,hakim62}. The interparticle dipole-dipole interactions (DDIs) are long-range and anisotropic, giving rise to behaviour such as magneto/electrostriction, geometric pattern formation and instabilities.  The latter include surface-rippling  in the presence of a perpendicular magnetic/electric field \cite{cowley67}, and is heralded by a roton-like softening of the dispersion relation at the ripple wavenumber.  The distinctive properties of ferrofluids are exploited in applications from tribology, to information display, and medicine \cite{rosensweig}. Recently, quantum ferrofluids have been realized in the form of superfluid Bose-Einstein condensates (BECs), composed of atoms with large magnetic dipole moments  \cite{Lahaye2009} such as $^{52}$Cr \cite{Griesmaier,beaufils08},  $^{164}$Dy \cite{Lu} and $^{168}$Er \cite{Aikawa}, and some signatures of FHD have been observed in these gases, including magnetostriction \cite{Lahaye2007} and {\it d}-wave collapse \cite{Lahaye2008}. Pattern formation \cite{parker09,ronen07,bohn09}, linked to roton-like excitations \cite{ronen07,ODell2003,Santos2003}, has been predicted but not yet seen.  Superfluidity of semiconductor microcavity polaritons, which are inherently dipolar, has also recently been demonstrated \cite{polaritons}.  

In this letter we consider the interplay between DDIs and vortices in a BEC. Vortices form the ``sinews and muscles" of fluids \cite{Kuchemann1965,Saffman1979}, and phenomena such as mixing processes, sunspots, tornadoes and synoptic scale weather phenomena are all driven by vortex dynamics \cite{Lugt1983}.  Vortices can also self-organize into large scale turbulent structures \cite{Saffman1979} and vortex crystals \cite{Aref2003}.  Even in pairs, vortices display rich behaviour, e.g. co-rotation of like-sign vortices, solitary-wave propagation of opposite-sign pairs, and reconnection and annihilation events \cite{Pismen}.  In a superfluid such as a BEC the vortices have a core of well-defined size and quantized vorticity.  Diverse vortex structures such as single vortices, vortex rings, pairs, lattices and turbulent states can be controllably generated \cite{Fetter} and imaged in real-time \cite{Freilich2010}.  In quasi-2D geometries, the paradigm of point vortices \cite{Aref1983,Middelkamp2011} can be realized.  2D Bose gases also provide a route to the Berezinsky-Kosterlitz-Thouless (BKT) transition \cite{BKT}, the thermal unbinding of vortex-antivortex pairs.  

Previous theoretical studies of a vortex in a 3D dipolar BEC found density ripples about the core \cite{pu,Wilson,abad} (related to the roton minimum mentioned above), an elliptical core \cite{pu}, and modified critical points for vortex nucleation \cite{ODell2007,Bijnen2007,abad,Ticknor2011,Kishor2012}. Here we consider the quasi-2D case and focus on the effective long-range and anisotropic potentials that are generated between entire vortices by DDIs.  We demonstrate the striking implications of these potentials on the motion of pairs of vortices.  Our results provide a microscopic model for understanding the role of DDIs in large-scale superfluid phenonema, such as the vortex lattice phases that have been predicted numerically \cite{lattices}, the BKT transition and quantum turbulence.

When the dipoles are aligned by an external field (as we shall assume throughout), the DDIs are described by,
\begin{equation}
U_{\rm dd}(\mathbf{r-r'})=\frac{C_{\rm dd}}{4\pi}\frac{1-3\cos^{2}\theta}{|\mathbf{r-r'}|^3},
\label{eq:DDI}
\end{equation}
where $\theta$ is the angle between the polarization direction and the inter-particle vector ${\bf r}-{\bf r'}$. 
For magnetic dipoles with moment $d$, $C_{\rm dd} = \mu_0 d^2$, where $\mu_0$  is the permeability of free space.  The recently made strongly dipolar $^{164}$Dy BEC  \cite{Lu} has $d=10 \mu_{\rm B}$ (Bohr magnetons).  The same interaction arises between polar molecules,  which can possess huge electric dipole moments and have been cooled close to degeneracy \cite{ni}.   A BEC with DDIs is described by the dipolar Gross-Pitaevskii equation (DGPE) \cite{Lahaye2009} in which the DDIs are incorporated via Eq.\ (\ref{eq:DDI}), and the isotropic van der Waals interactions (VDWIs) via a local pseudopotential $U_{\rm vdw}({\bf r}-{\bf r}')=g_{\rm 3D} \delta(\mathbf{r}-\mathbf{r'})$  \cite{Pethick2002}. The relative strength of the DDIs is parameterized via the ratio $\varepsilon_{\rm dd}= C_{\rm dd}/3g_{\rm 3D} $ \cite{Lahaye2009}.  This parameter has a natural value $\varepsilon_{\rm dd} \lappeq 1$.  However, the ability to tune $g_{\rm 3D}$ between $-\infty$ and $+\infty$ via Feshbach resonance has enabled the realization of a purely dipolar gas ($\varepsilon_{\rm dd}=\infty$) \cite{koch}.  Further, it is predicted that $C_{\rm dd}$ can be reduced below its natural, positive value, including to negative values, by external field rotation \cite{giovanazzi}.  As such, a large parameter space $-\infty <\varepsilon_{\rm dd} < \infty$, with positive or negative $g_{\rm 3D}$ and $C_{\rm dd}$, is possible.

We consider bosonic dipoles of mass $m$, free in the transverse $(\boldsymbol{\rho})$ plane and with harmonic trapping in the axial $(z)$ direction.  The axial trap frequency $\omega_z$ is sufficiently strong that the BEC is frozen into the axial ground harmonic state of width $l_z=\sqrt{\hbar/m \omega_z}$ \cite{Parker}.  The polarization axis is at angle $\alpha$ to the $z$-axis in the $xz$ plane.  Integrating over $z$ gives the effective 2D DGPE for the 2D wave function $\psi(\boldsymbol{\rho},t)$ \cite{Ticknor2011},
\begin{equation}
i \hbar \frac{\partial \psi}{\partial t}=\left [-\frac{\hbar^2}{2m}\nabla^2+g|\psi|^2+\Phi - \mu\right ]\psi,
\label{eqn:dgpe}
\end{equation}
with chemical potential $\mu$. The mean-field potentials  $g\vert \psi(\boldsymbol{\rho},t) \vert^{2}$ and $\Phi(\boldsymbol{\rho},t)=\int U^{\rm 2D}_{\rm dd}(\boldsymbol{\rho-\rho}')n(\boldsymbol{\rho'},t)~d\boldsymbol{\rho}'$ account for the VDWIs and DDIs, respectively, where $n=|\psi|^2$ is the 2D particle density, $U^{\rm 2D}_{\rm dd}$ is the effective 2D DDI potential,  and $g=g_{\rm 3D}/\sqrt{2\pi}l_z$. While an explicit expression for $U^{\rm 2D}_{\rm dd}(\boldsymbol{\rho})$ exists \cite{Bao2010}, it is convenient to work in Fourier space and use the convolution theorem $\Phi(\boldsymbol{\rho},t)=\mathcal{F}^{-1} \left[\tilde{U}^{\rm 2D}_{\rm dd}({\bf k}) \tilde{n}({\bf k},t)\right]$ \cite{Fischer2006,Ticknor2011}.   The  Fourier transform of $U^{\rm 2D}_{\rm dd}(\boldsymbol{\rho})$ is,
\begin{equation}
\tilde{U}^{\rm 2D}_{\rm dd}(\mathbf{q})=\frac{4\pi g_{\rm dd}}{3} \left[F_{\parallel}\left (\mathbf{q} \right ) \sin^2\alpha +F_{\perp}\left (\mathbf{q} \right )\cos^2\alpha\right],
\end{equation}
with $\mathbf{q}=\mathbf{k}l_z/\sqrt{2}$, $F_{\parallel}({\bf q})=-1+3\sqrt{\pi}\frac{q_{x}^{2}}{q}e^{q^2} \mathrm{erfc}(q)$, $F_{\perp}({\bf q})=2-3\sqrt{\pi}q e^{q^2} \mathrm{erfc}(q)$, and   $g_{\rm dd}=C_{\rm dd}/3 \sqrt{2 \pi}l_z$.  It follows that $\varepsilon_{\rm dd}= g_{\rm dd}/g$.

\begin{figure}[b]
	\includegraphics[width=1\columnwidth,angle=0]{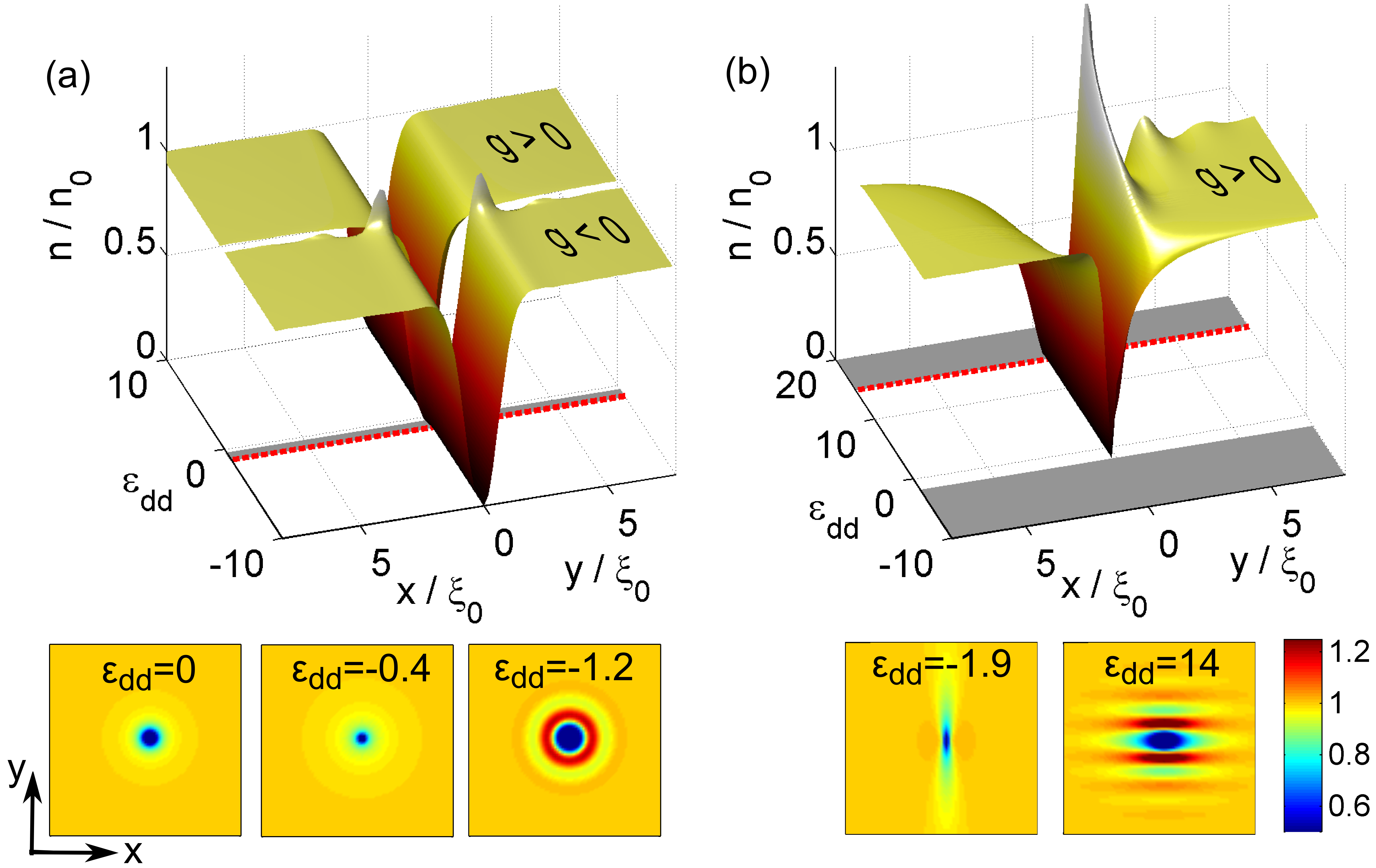}
		\caption{Vortex density profiles in the presence of DDIs parameterized by $\varepsilon_{\rm dd}=g_{\rm dd}/g$. The left (right) side of each figure shows the profile along $x$ ($y$). (a) Polarization along $z$ ($\alpha=0$).    Stable density ripples form close to the onset of the RI (dashed red line).  (b) For off-axis dipoles ($\alpha=\pi/4$) the vortex becomes highly anisotropic.  Insets: vortex density $n(x,y)$ for examples of $\varepsilon_{\rm dd}$ [area $(40\xi_0)^2$ for each].  Grey bands indicate unstable regimes where no steady state solutions exist, while the labels $g>0$ and $g<0$ on the individual solution sheets indicate that they are only stable for the specified value of $g$.}
		\label{fig:single_vortices}
\end{figure}

A homogeneous 2D dipolar gas of density $n_0$ has uniform dipolar potential $\Phi_0=n_0 g_{\rm dd}(3 \cos^2 \alpha -1)$ and chemical potential $\mu_0=n_0 \left(g+g_{\rm dd} [3\cos^2\alpha-1]\right )$ \cite{Ticknor2011,Parker,Bao2010}.    At the ``magic angle'' $\alpha_0=\textrm{arccos}\left(1/\sqrt{3}\right)\approx 54.7^\circ$, $\Phi_0$ is zero.  For $\alpha< \alpha_0$  ~$\Phi_0$ is net repulsive (attractive) for $g_{\rm dd}>0$ ($<0$), while for $\alpha > \alpha_0$ it is net attractive (repulsive) for $g_{\rm dd}>0$ ($<0$).  
The homogeneous system suffers two key instabilities: the phonon and roton instabilities.   The phonon instability (PI), familiar from conventional BECs \cite{Pethick2002}, is associated with unstable growth of zero momentum modes.  It arises when the net local interactions become attractive, i.e. when $\mu_0 < 0$.  In terms of $\varepsilon_{\rm dd}$, it follows  that the PI arises for $\varepsilon_{\rm dd}< [1- 3 \cos^2 \alpha ]^{-1}$ when $g>0$ or $\varepsilon_{\rm dd}> [1- 3 \cos^2 \alpha ]^{-1}$ when $g<0$ \cite{Santos2003,ODell2004,Fischer2006,Lahaye2009,Bijnen}.

The roton instability (RI) is associated with unstable growth of finite momentum modes.  DDIs can induce a roton dip at finite momentum in the excitation spectrum \cite{ODell2003,Santos2003,Wilson} which, for certain parameters, softens to zero energy.  In 3D a collapse then ensues via density ripples with wave fronts aligned either along the polarization axis (for $C_{\rm dd}>0)$ or perpendicular to it (for $C_{\rm dd}<0$) due to the preferred alignment of dipoles end-to-end  or side-by-side, respectively  \cite{parker09}.  Close to the RI stable density ripples arise when the roton mode mixes into the ground state \cite{pu,Wilson}.  Typically, the RI is induced by the attractive part of the DDI. An exception arises for the 2D gas with dipoles polarized along $z$; the attractive part of the DDI lies out of the plane where the particles cannot access it \cite{Fischer2006} but a roton can be induced via attractive local interactions \cite{Klawunn2009}.  For dipoles polarized in the plane, the particles can access the attractive part of the DDI, and the `conventional' dipolar roton is supported.

Understanding the vortex-vortex interaction requires first understanding the single vortex solutions: While DDIs have been shown to modify the vortex solution within the context of a trapped 3D BEC, the case of 2D and homogeneous systems relevant to us has not been explored.  We obtain vortex solutions and dynamics by numerically solving Eq.~(\ref{eqn:dgpe}) \cite{SM}.  Density, energy and length are scaled in units of $n_0$, $\mu_0$ and the corresponding healing length $\xi_0=\hbar/\sqrt{m \mu_0}$, respectively.  The vortex core size is of the order of $\xi_0$, which diverges as $\mu_0 \rightarrow 0$.  The 2D approximation is valid for $\sigma=l_z/\xi_0 \leq 1$; we choose $\sigma=0.5$.  We consider the representative cases of $\alpha=0$ (dipoles parallel to the $z$-axis) and $\alpha=\pi/4$ (dipoles tipped partly along $x$-axis):

\underline{$\alpha=0$:}   Figure~\ref{fig:single_vortices}(a) shows the vortex density along $x$ and $y$ as a function of $\varepsilon_{\rm dd}$.  The dipolar potential, and hence density profile, are axi-symmetric.  For $\varepsilon_{\rm dd}=0$ (left inset) the vortex has the standard axi-symmetric core of vanishing density of width $\xi_0$ \cite{Pethick2002}.  For $g>0$ the system is stable from $\varepsilon_{\rm dd}=-0.5$, the PI threshold, upwards.  No RI is observed for $g>0$, as expected \cite{Fischer2006}.  Meanwhile, for $g<0$ we only find solutions for $\varepsilon_{\rm dd} \lappeq -1.16$, the RI threshold (red dashed line).   

The vortex structure for $\varepsilon_{\rm dd} \neq 0$ is almost identical to the $\varepsilon_{\rm dd}=0$ case \cite{HL}, apart from in two regimes.  As one approaches $\varepsilon_{\rm dd}=-0.5$ from above, the vortex core becomes increasingly narrow with respect to $\xi_0$ (middle inset), due to the cancellation of explicit contact interactions.  Meanwhile, as the RI is approached from below, axi-symmetric density ripples appear around the vortex (like in the 3D case \cite{pu,Wilson,abad}), decaying with distance and with an amplitude up to $\sim 20\% \, n_0$ (see third inset).   The ripple wavelength, which reflects the roton wavelength, is $\approx 4 \xi_0$ \cite{wavelength} implying that our treatment is self-consistently 2D because $\sigma=l_{z}/\xi_{0}=0.5$.

\underline{ $\alpha=\pi/4$:} The axi-symmetry of the dipolar potential and density is lost [Fig.~\ref{fig:single_vortices}(b)].  For $g_{\rm dd}>0$ ($<0$) the dipoles lie preferentially along $x$ ($y$).  
For $g<0$ we find no stable solutions (down to $\varepsilon_{\rm dd}=-20$) due to the RI.  For $g>0$ we observe the RI, unlike for $\alpha=0$, since the atoms now feel the attractive part of $U_{\rm dd}$.  The RI occurs for $\varepsilon_{\rm dd}\gappeq 14.9$ (red dashed line).  Below this, stable solutions exist down to the PI threshold $\varepsilon_{\rm dd}=-2$.   The vortex core is elongated along $x$ for $\varepsilon_{\rm dd}>0$, and reverses for $\varepsilon_{\rm dd}<0$.   Close to the RI, density ripples form about the vortex (of large amplitude up to $\sim 40\%$) aligned along $x$ due to the preferred end-to-end alignment of dipoles along $x$.    

We see the same qualitative behaviour for all $0<\alpha < \alpha_0$, albeit with shifted RI and PI thresholds.  For cases with $\alpha > \alpha_0$, we see a reversal in the $\varepsilon_{\rm dd}$ dependence, with no solutions for $g>0$ and ripples polarized along $y$.

\begin{figure}[t]
	\includegraphics[width=1\columnwidth]{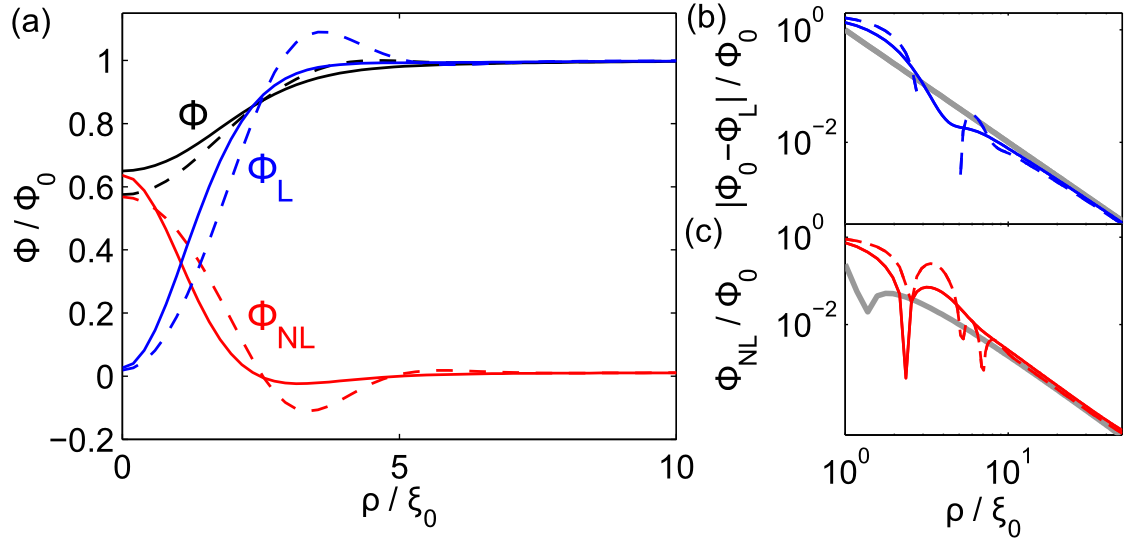}
	\caption{(a) The total dipolar potential $\Phi$ (black), local component $\Phi_{\rm L}$ (blue) and non-local component $\Phi_{\rm NL}$ (red) for a vortex with ripples ($\varepsilon_{\rm dd}=-1.2$, $g<0$, dashed line) and without ripples ($\varepsilon_{\rm dd}=5$, $g>0$, solid line).  (b) The decay of $\Phi_{\rm L}/\Phi_0$, compared with $1/\rho'^2$ (grey line).  (c) The decay of $\Phi_{\rm NL}/\Phi_0$, compared with $(A\ln \rho'+B)\sigma/\rho'^3$. Parameters: $\alpha=0$, $\sigma=0.5$.}
	\label{fig:vortex_dipolar_potential}
\end{figure}

From the FHD perspective, the depleted density due to a vortex acts like a lump of ``anti-dipoles" whose charges have been reversed \cite{klawunn}.  Let us calculate the mean-field dipolar potential $\Phi (\boldsymbol{\rho})$ generated by such a defect located at $\boldsymbol{\rho}=0$; we shall see shortly how it modifies the vortex-vortex interaction.  The total dipolar energy is $E_{\mathrm{dd}}=\tfrac{1}{2}\int \Phi(\boldsymbol{\rho}) n(\boldsymbol{\rho}) d \boldsymbol{\rho}$ so $\Phi(\boldsymbol{\rho})$ plays a role analogous to an electrostatic potential with the density of dipoles $n(\boldsymbol{\rho})$ analogous to the charge density.
For simplicity we consider $\alpha=0$  and the illustrative cases of $\varepsilon_{\rm dd}=-1.2$ (vortex with ripples, $g<0$) and $5$ (vortex with no ripples, $g>0$) [Fig.~\ref{fig:vortex_dipolar_potential}(a)].  As $\rho \rightarrow \infty$, $\Phi \rightarrow  \Phi_0 = n_0 g_{\rm dd}(3\cos^2 \alpha -1)$, the homogeneous result, while  for $\rho \lappeq 5\xi$, $\Phi$ is dominated by the core structure and ripples (where present).   It is insightful to consider $\Phi$ as the sum of a local term $\Phi_{\rm L}(\boldsymbol{\rho})=n(\boldsymbol{\rho}) g_{\rm dd}(3 \cos^2 \alpha -1)$ and a non-local term $\Phi_{\rm NL}(\boldsymbol{\rho})$ \cite{ODell2004,Parker,Bao2010}.   The latter is generated by variations in the density \cite{ODell2004,Parker} and vanishes for a homogeneous system.  The generic functional form of $\Phi$ at long-range is revealed by assuming the vortex ansatz $n(\rho)=n_0[1- 1/(1+\rho'^2)]$, where $\rho'=\rho/\xi_0$ \cite{ODell2007,Fetter_vortex}.  We employ a first-order expansion in $\sigma$ of $\tilde{U}^{\rm 2D}_{\rm dd}(\mathbf{k}) = g_{\rm dd}\left(2-\sqrt{\frac{9 \pi }{2}} k \sigma \right) +\mathcal{O}\left(\sigma ^2\right)$ and the Hankel transform $\tilde{n}(\mathbf{k})=n_0\delta (k)/k+n_0 K_0( k / \sqrt{2}) / 2$,
where $K_0(k)$ is a Bessel function of the 3rd kind \cite{Abramowitz}. Then, to first order in $\sigma$ and third order in $1/\rho'$,
\begin{alignat}{1}
\frac{\Phi(\mathbf{\rho})}{\Phi_0}\sim \left(1 -\frac{1}{\rho'^{2}}\right)+\left(\frac{A \ln \rho' + B}{\rho'^{3}}\right) \sigma,\label{eq:analytic_phidd}
\end{alignat}
with constants $A=-\sqrt{9 \pi/8}\approx -1.88$ and $B=(\ln 2 -1)A\approx 0.577$.   The first term corresponds to the {\em local} contribution of the vortex dipolar potential [Fig.~\ref{fig:vortex_dipolar_potential}(b), grey line].  It physically arises from the $1/\rho'^2$ decay of vortex density at long-range, and is also present for non-dipolar vortices (albeit controlled by $g$ and not $g_{\rm dd}$) \cite{Wu1994}.   The second term describes the {\em non-local} contribution [Fig.~\ref{fig:vortex_dipolar_potential}(c), grey line].  This vanishes in the true 2D limit $\sigma=0$ since the volume of anti-dipoles in the core vanishes.  Equation (\ref{eq:analytic_phidd}) agrees with numerical calculations in the limit $\rho^{\prime}\gg 1$ [Fig 2(b,c)]. 
The dominant scaling of $\Phi_{\rm NL}$ as $\ln \rho'/\rho'^3$, and not $1/\rho'^3$, shows us that the vortex does not strictly behave as a point-like collection of dipoles at long-range.  This is due to the slow, power-law recovery of the vortex density to $n_0$.  We have checked that exponentially-decaying density profiles, e.g.\ $ \tanh^2 (\rho')$, do lead to a $1/\rho'^3$ scaling of $\Phi_{\rm NL}$ at long-range.  For $\alpha \neq 0$, $\Phi$ also varies anisotropically as $\cos^2 \theta$ at long-range.

We now explore the vortex-vortex interaction through the interaction energy \cite{SM}.  
\begin{figure}[b]
	\includegraphics[width=1\columnwidth]{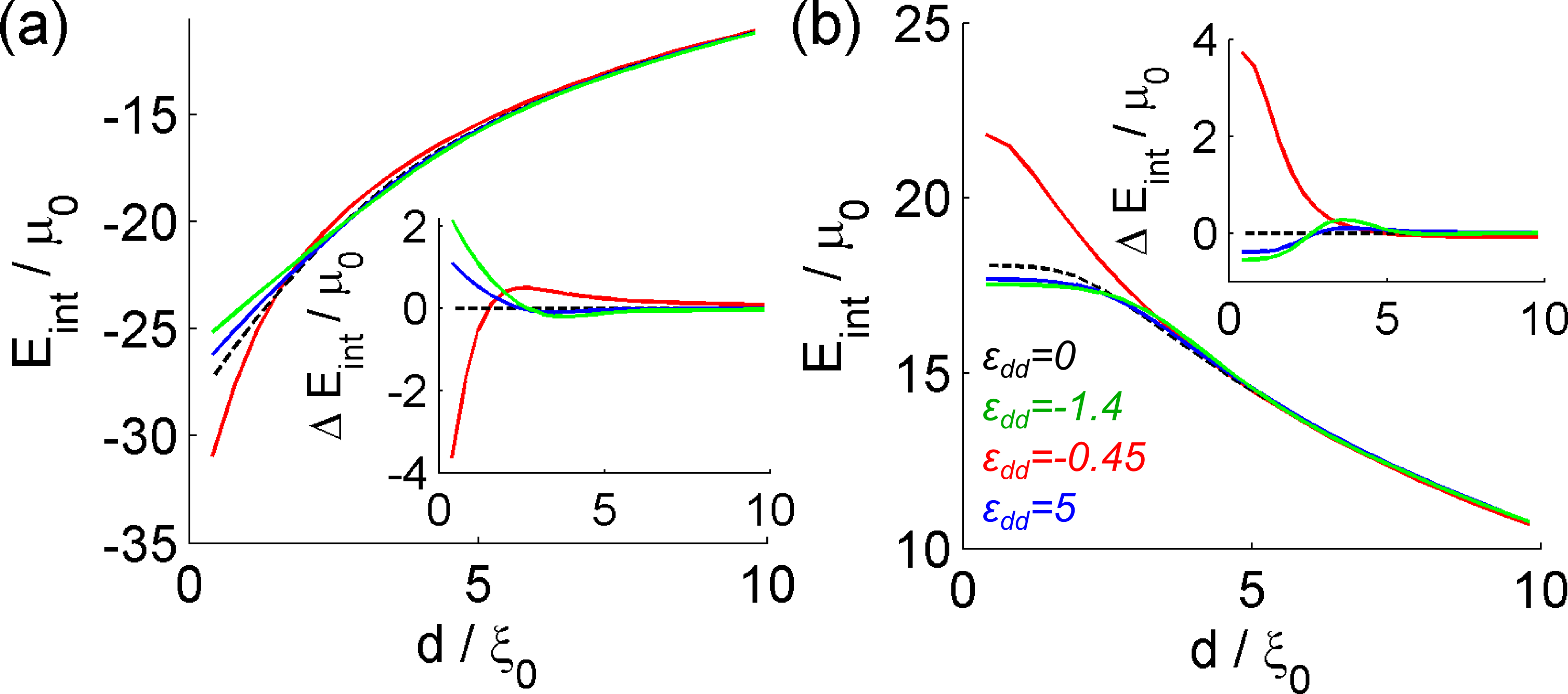}
	\caption{Vortex interaction energy $E_{\rm int}$ versus separation for (a) VA and (b) VV pairs with various $\varepsilon_{\rm dd}$ values.   Insets: energy difference from the non-dipolar case, $\Delta E_{\rm int}=E_{\rm int}(d,\varepsilon_{\rm dd})-E_{\rm int}(d,\varepsilon_{\rm dd}=0)$.  Parameters: $\alpha=0$, $\sigma=0.5$.
}
\label{fig:ints}
\end{figure}
Figure~\ref{fig:ints} shows $E_{\rm int}$ for $\alpha=0$ and (a) vortex-antivortex (VA) and (b) vortex-vortex (VV) pairs as a function of their separation $d$.  For the conventional $\varepsilon_{\rm dd}=0$ case (dashed line) $E_{\rm int}$ is negative (positive) for VA (VV) pairs due to the cancellation (reinforcement) of velocity fields at large distance.  For $d \gappeq 4\xi_0$, $E_{\rm int}(d)\approx (2\pi q_1 q_2 \hbar^2 n_0 / m) \ln (R/d)$, the hydrodynamic (coreless vortex) prediction \cite{Pethick2002} for system size $R \gg d$.  For $d \lappeq 4\xi_0$, the cores overlap causing $|E_{\rm int}|$ to flatten off.

When $\varepsilon_{\rm dd} \neq 0$ we still find that $E_{\rm int}(d)$ 
approaches the hydrodynamic result at large $d$. Although the dominant 
$1/\rho^2$ term in $\Phi$ evaluates to $E_{\rm dd} \approx - \pi n_{0} 
\Phi_{0} N_{\rm v} \ln(\xi_{0}/R)$, where $N_{\rm v}$ is the number of 
vortices, this does not depend upon $d$ for large separations since 
$\Phi$ arises from vortex density profiles. These do not show any 
special behavior at $\rho=d$, unlike superposed velocity fields which 
change from cancellation to reinforcement or vice versa.  Elsewhere, significant deviations to $E_{\rm int}(d)$ arise, particularly at short-range $d \lappeq 3\xi_0$.  This deviation is most striking for $-0.5<\varepsilon_{\rm dd}<0$ ($g>0$), e.g.\ $\varepsilon_{\rm dd}=-0.45$ (red lines), for which $|E_{\rm int}|$ increases dramatically as $d$ decreases.  This is due to the narrowing of the vortex core (relative to $\xi_0$) as $\varepsilon_{\rm dd} \rightarrow -0.5$, reducing the core overlap as $d \rightarrow 0$.  Outside of this range, for small $d$ ($d \lappeq 3\xi_0$), DDIs reduce $|E_{\rm int}|$.  For values of $\varepsilon_{\rm dd}$ that support density ripples, $|E_{\rm int}|$ features a small peak at $d \sim 4\xi$ due to ripple overlap.  Further out, $E_{\rm int}$ decays as $1/\rho^2$ towards the $\varepsilon_{\rm dd}=0$ result.   Note that as $|\varepsilon_{\rm dd}|$ is increased, $E_{\rm int}(d)$ tends towards a fixed behaviour, independent of the sign of $\varepsilon_{\rm dd}$ (since both cases become dominated by large, positive $g_{\rm dd}$).

\begin{figure}
	\includegraphics[width=1\columnwidth]{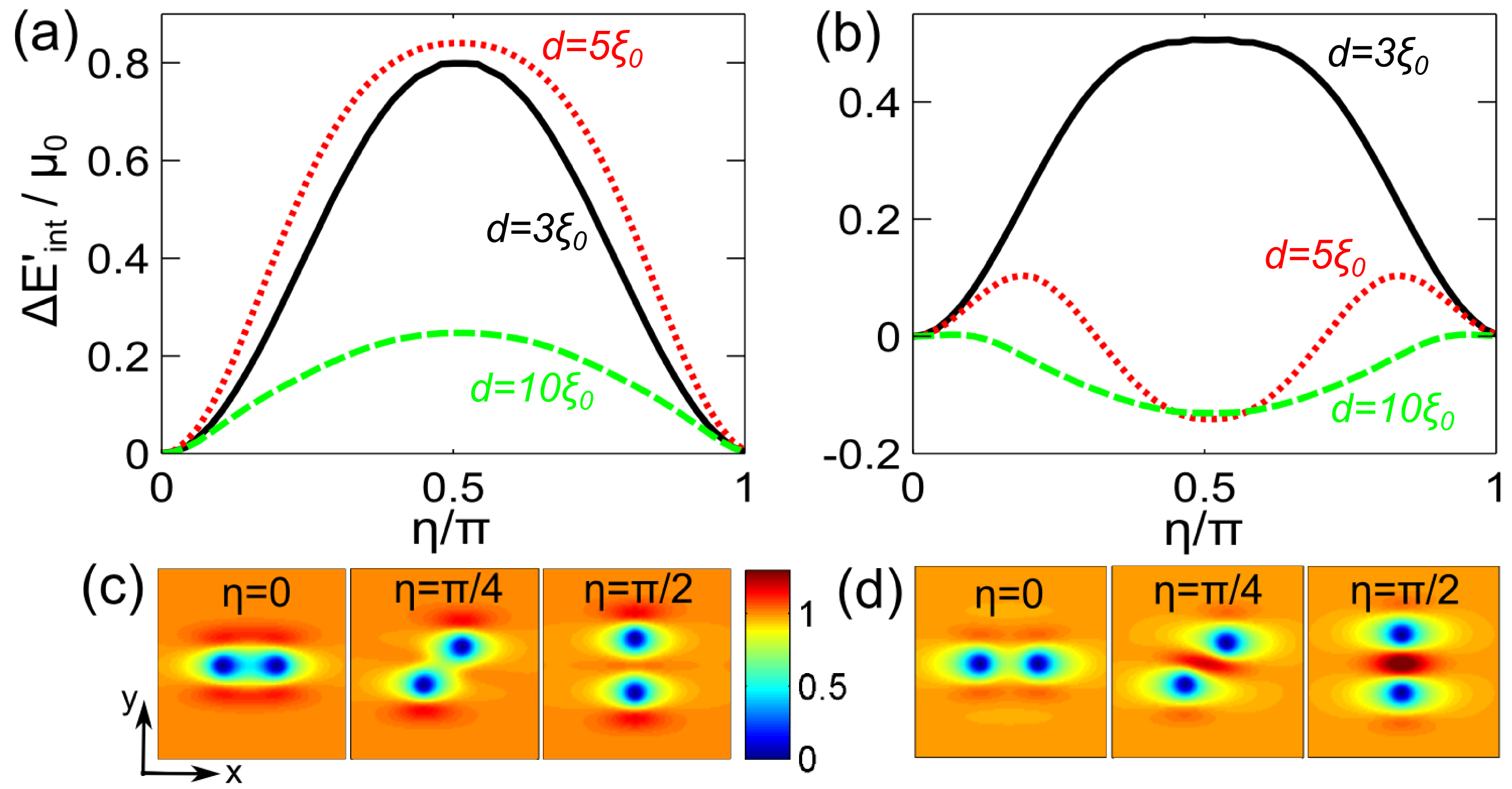}
	\caption{Angular dependence of the vortex interaction energy $\Delta E'_{\rm int}(\eta)=E_{\rm int}(\eta)-E_{\rm int}(\eta=0)$, for (a) VA and (b) VV pairs of differing separation.  Parameters: $\alpha=\pi/4$, $\sigma=0.5$ and $\epsilon_{\rm dd}=5$ ($g>0$).   Example (c) VA and (d) VV pair density profiles for $d=5\xi_0$ [area $(20\xi_0)^2$ for each]. }
\label{fig:angle}
\end{figure}

\begin{figure}[b]
\centering
	\includegraphics[width=\columnwidth]{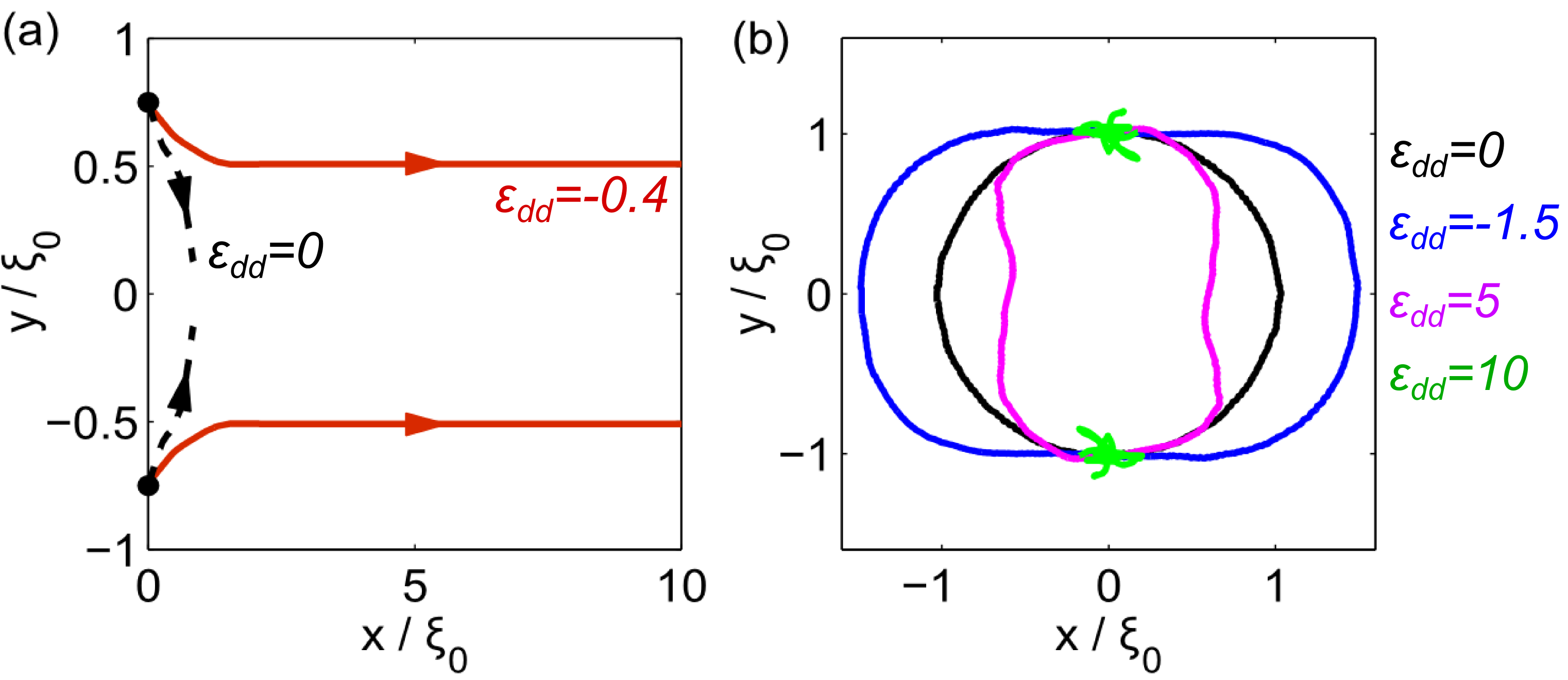}
	\caption{(a) A VA pair (initial separation $d=1.5\xi$) with no DDIs annihilates while DDIs ($\alpha=0$) stabilize the pair (red lines).  (b) A VV pair ($d=2\xi_0$) with no DDIs co-rotates in a circular path.  Off-axis ($\alpha=\pi/4$) DDIs lead to anisotropic paths and even suppression of co-rotation altogether.  }
\label{fig:pairs}
\end{figure}

For $\alpha \neq 0$, $E_{\rm int}$ depends on orientation angle of the pair $\eta$, defined as the angle between the line joining the vortices and the  in-plane polarization direction.  We illustrate this in Fig.\ \ref{fig:angle} for $\alpha=\pi/4$ and $\varepsilon_{\rm dd}=5$.   For (a) the VA pair, $E_{\rm int}$ is maximal for $\eta=\pi/2$ and decreases monotonically to 0 as $\eta \rightarrow 0, \pi$.  For moderate separations $d \lappeq 6 \xi_0$ (dotted red line/solid black line), the vortex ripples dominate the vortex interaction (see inset).   For $\eta=0$, the ripples from each vortex (aligned in $x$) merge side-by-side.  As $\eta$ is increased, this side-by-side formation is broken at energetic cost. For $\eta \sim \pi/2$ one might expect, for appropriate $d$, a scenario where the inner ripples from each vortex overlap at energetic benefit. However, the inter-vortex density is suppressed due to the high flow velocity there (Bernoulli's principle).  Beyond the ripples, $E_{\rm int}$ approaches a sinusoidal dependence on $\eta$ (dashed green line). 

For (b) the VV pair no such suppression occurs.  As the pair is rotated, and for suitably large $d$ ($d \gappeq 2\xi_0$), the inner ripples combine at significant energetic benefit such that $E_{\rm int}$ is minimal for $\eta=\pi/2$.   At smaller separations, this overlap cannot occur and $E_{\rm int}(\eta)$ is dominated by the effect of the outer ripples (as for the VA pair), such that the $\eta=0$ case is most energetically favoured.  For intermediate $d$ ($d \sim 3\xi_0$), there is energetic competition between the inner and outer ripples.  At larger $d$, the modulation becomes approximately sinusoidal with $\eta$.

The effects of DDIs upon vortex pair dynamics are now considered.  For $\varepsilon_{\rm dd}=0$, VA pairs form moving solitary wave solutions for $d \gappeq 2\xi_0$ \cite{pairs}.  For $d \lappeq 2\xi_0$ the cores overlap and the pair is unstable to annihilation [Fig.~\ref{fig:pairs}(a)].  In Fig.~\ref{fig:single_vortices}(a), we saw significant core narrowing (relative to $\xi_0$) for $\varepsilon_{\rm dd}\approx -0.4$.  These conditions can stabilize the VA pair against annihilation (Fig. ~\ref{fig:pairs}(a), red lines) \cite{transient}. 
For $\varepsilon_{\rm dd}=0$ (or $\alpha=0$), VV pairs undergo circular co-rotation \cite{Pismen} (black line curve, Fig. \ref{fig:pairs}(b)).  Since in-plane ($\alpha \ne 0$) DDIs cause $E_{\rm int}$ to vary as the VV pair rotates, the vortices co-rotate in an anisotropic path [blue and magenta curves, Fig. \ref{fig:pairs}(b)].  Moreover, with strong anisotropic ripples for small separations, the vortices are unable to co-rotate, instead ``wobbling" about their initial positions [green curves, Fig. \ref{fig:pairs}(b)].

We have shown that the interaction of two vortices can be significantly different in quantum ferrofluids than in conventional superfluids.  At short-range the vortex-vortex interaction is strongly modified by the changed shape and peripheral density ripples of each vortex.   At longer range, each vortex experiences the dipolar mean-field potential of the other, with $1/\rho^2$ and $\ln(\rho)/\rho^3$ contributions.  The vortex-vortex interaction is most significantly modified up to mid-range separations ($d \lappeq 10\xi$), beyond which it reduces to the usual hydrodynamic behaviour.  When the dipoles have a component in the plane, the vortex-vortex interaction becomes anisotropic.  

The vortex-vortex interaction is a pivotal building block for understanding macroscopic superfluid phenomena.  For example, it is the key input parameter for vortex line-based modelling of quantum turbulence \cite{carlo}, evaluating the energetics of vortex pair unbinding central to the BKT transition \cite{BKT}, and vortex point-based modelling of vortex crystals \cite{Aref2003}.  The striking effects of the dipolar interactions we have shown for the dynamics of vortex pairs, i.e.  modification of the vortex-antivortex annihilation threshold and anisotropic co-rotation of vortex-vortex pairs, suggest that dipolar interactions will open up interesting new regimes in such macroscopic systems of superfluid vortices.

BM acknowledges support from the Overseas Research Experience Scholarship, University of Melbourne, and DO acknowledges support from NSERC (Canada).


\begin{thebibliography}{10}
\bibitem{HL} This is our motivation for scaling length in terms of $\xi_0$.
\bibitem{transient} The initial motion of the vortices towards each other is a transient effect associated with starting the simulation with a vortex pair solution obtained in the static frame.
\end{thebibliography}
\end{document}